\documentclass[12pt]{article}
\begin{document}
\setcounter{page}{1}
\def\theequation{\arabic{section}.\arabic{equation}}
\def\theequation{\thesection.\arabic{equation}}
\setcounter{section}{0}

\title{~\rightline{\normalsize {\it If you can't join 'em, beat
'em}}\\[-0.1in] ~\rightline{\normalsize {\it Julian
Schwinger}}\\[0.2in] Dynamics of low--energy nuclear forces and Solar
Neutrino Problems in the Nambu--Jona--Lasinio model of light nuclei}

\author{A. N. Ivanov~\thanks{E--mail: ivanov@kph.tuwien.ac.at, Tel.:
+43--1--58801--14261, Fax: +43--1--58801--14299}~${
^\ddagger}$, H. Oberhummer~\thanks{E--mail: ohu@kph.tuwien.ac.at,
Tel.: +43--1--58801--14251, Fax: +43--1--58801--14299} ,
N. I. Troitskaya~\thanks{Permanent Address: State Technical
University, Department of Nuclear Physics, 195251 St. Petersburg,
Russian Federation} , M. Faber~\thanks{E--mail:
faber@kph.tuwien.ac.at, Tel.: +43--1--58801--14261, Fax:
+43--1--58801--14299}}

\date{\today}

\maketitle

\begin{center}
{\it Institut f\"ur Kernphysik, Technische Universit\"at Wien,\\
Wiedner Hauptstr. 8-10, A-1040 Vienna, Austria}
\end{center}

\begin{center}
\begin{abstract}
The Solar Neutrino Problems (SNP's) are analysed in the
Nambu--Jona--Lasinio model of light nuclei. In this model a possible
clue to the solution of the SNP's is in the reduction of the solar
neutrino fluxes relative to the predicted by the Standard Solar Model
through the decrease of the solar core temperature.  The former can be
realized through the enhancement of the astrophysical factor for the
solar proton burning. The enhancement the upper bound of which is
restricted by the helioseismological data goes dynamically via the
contribution of the nucleon tensor current coupled to the deuteron.
The agreement of the reduced solar neutrino fluxes with the
experimental data can be reached within a scenario of vacuum
two--flavour neutrino oscillations without a fine tuning of the
neutrino--flavour oscillation parameters. In the Nambu--Jona--Lasinio
model of light nuclei an enhancement of the astrophysical factor for
the solar proton burning entails a change of the cross sections for
neutrino and anti--neutrino disintegration of the deuteron at low
energies. This provides a theoretical foundation for a new check of a
value of the astrophysical factor in terrestrial laboratories.
\end{abstract}
\end{center}

\begin{center}
PACS: 11.10.Ef, 13.75.Cs, 14.20.Dh, 21.30.Fe, 26.65.+t\\
\noindent Keywords: deuteron, proton burning, solar neutrino fluxes
\end{center}

\newpage

\section{Introduction}
\setcounter{equation}{0} 

\hspace{0.2in} The weak nuclear reaction p + p $\to$ D + e$^+$ +
$\nu_{\rm e}$, the solar proton burning, plays an important role in
Astrophysics [1,2]. It gives start for the p--p chain of
nucleosynthesis in the Sun and main--sequence stars [1,2]. In the
Standard Solar Model (SSM) [3] the total (or bolometric) luminosity of
the Sun $L_{\odot} = (3.846\pm 0.008)\times 10^{26}\,{\rm W}$ is
normalized to the astrophysical factor $S_{\rm pp}(0)$ for the solar
proton burning. The recommended value $S^{\rm SSM}_{\rm pp}(0) =
4.00\times 10^{-25}\,{\rm MeV b}$ [4] has been found by averaging over
the results obtained in the Potential model approach (PMA) [5,6] and
the Effective Field Theory (EFT) approach [7,8]. As has been shown
recently [9] {\it the inverse and forward helioseismological approach
indicate that higher values of $S_{\rm pp}(0)$ seem more
favoured}. However, as has been stated by Degl'Innocenti, Fiorentini
and Ricci [10] the helioseismological data restrict the value of the
astrophysical factor $S_{\rm pp}(0)$ and predict $0.94 \le S_{\rm
pp}(0)/S^{\rm SSM}_{\rm pp}(0) \le 1.18$.

The value of the astrophysical factor $S_{\rm pp}(0)$ for the solar
proton burning is very closely connected with the Solar Neutrino
Problem formulated by Bahcall [1] as a discrepancy between theoretical
and experimental values for the solar neutrino fluxes of high energy
neutrinos.  Recently [11] following the contemporary experimental data
GALLEX[12], SAGE [13], HOMESTAKE [14], KAMIOKANDE [15] and
SUPERKAMIOKANDE [16] Bahcall formulated three Solar Neutrino Problems
(SNP's), {\it three different discrepancies between the calculations
and the observations} of solar neutrino fluxes. In Table 1 we adduce
the experimental data on the solar neutrino fluxes, whereas Table
2 contains the theoretical values of the solar neutrino fluxes
calculated within the SSM [3] and normalized to the recommended value
for the astrophysical factor $S^{\rm SSM}_{\rm pp}(0)$ [4]. According
to Bahcall's classification [11]:

The {\it first} SNP is the disagreement between the calculations and
observations for the chlorine experiment [14]. The measured rate is
$2.56\pm 0.23\,{\rm SNU}$\,\footnote{Everywhere in the paper we use
statistical and systematic experimental errours combined in quadrature
[11] $\sigma =\sqrt{({\rm stat.})^2 + ({\rm syst.})^2}$}, whereas the
theoretical prediction is about 3 times larger
$7.7^{+1.2}_{-1.1}\,{\rm SNU}$ As has been emphasized by Bahcall [11]
{\it most of the predicted rate in the chlorine experiment is from the
rare, high energy ${^8}{\rm B}$ neutrinos, although the ${^7}{\rm Be}$
neutrinos are expected to contribute significantly}.

The {\it second} SNP results from a comparison of the measured event
rates in the chlorine experiment and in the Japanese purewater
experiments, these are KAMIOKANDE [15] and SUPERKAMIOKANDE
[16]. According to the SSM [3] the main contribution to neutrino
fluxes measured by KAMIOKANDE and SUPERKAMIOKANDE Collaborations comes
from the $\beta$ decay of ${^8}{\rm B}$, ${^8}{\rm B} \to {^8}{\rm
Be^*}$ + e$^+$ + $\nu_{\rm e}$, in the solar core.  There is also a
contribution from the {\it hep} reaction, p + ${^3}{\rm He}$ $\to$
${^4}{\rm He}$ + e$^+$ + $\nu_{\rm e}$ [11].

The {\it third} SNP is related to the gallium experiments, GALLEX and
SAGE. Formally the experimental rates measured by these groups
evidence the exclusion of all contributions save the pp neutrinos
produced in the reaction p + p $\to$ D + e$^+$ + $\nu_{\rm e}$.

As has been stated by Bahcall [11] the experimental data obtained by
all five solar neutrino experimental groups, GALLEX, SAGE, HOMESTAKE,
KAMIOKANDE and SUPERKAMIOKANDE, can be fitted well within the
approaches involving neutrino flavour oscillations, either vacuum
oscillations suggested by Gribov and Pontecorvo [17--20] or resonant
matter oscillations suggested by Wolfenstein, Mikheyev and Smirnov
[21] so--called MSW effect [22]. The only forthcoming of these fits is
in the necessity to make a very fine tuning of neutrino--flavour
oscillation parameters: 1) the squared mass differences $\Delta
m^2_{ij} = m^2_i - m^2_j$, the mixing angles $\theta_{ij}$ and so on
where the indices run over the number of oscillating neutrino flavours
$i({\rm or}\,j) = \nu_{\rm e}, \nu_{\mu}, \nu_{\tau},\ldots $.
 
An alternative way that does not demand a fine tuning of the
neutrino--flavour oscillation parameters and leads to the solution of
the SNP's can go, for example, through the reduction of the solar
neutrino fluxes in the solar core caused by the decrease of the solar
core temperature. The former can be related to the enhancement of the
astrophysical factor for the solar proton burning.  If one would
follow the SSM such an enhancement of the astrophysical factor should
be restricted by the helioseismological data [10]. This constraint
prohibits the possibility to reduce at once the theoretical values of
the solar neutrino fluxes in agreement with experimental data. This
implies that a secondary reduction is required. Such a secondary
reduction of the theoretical solar neutrino fluxes but taking place
already outside the solar core can be induced, for example, by
neutrino--flavour oscillations. After the reduction of the solar
neutrino fluxes in the solar core the final result, most likely,
should not be sensitive to the mechanism of neutrino--flavour
oscillations.  Therefore, the secondary reduction of the theoretical
solar neutrino fluxes can be carried out, for example, within a vacuum
two--flavour neutrino oscillation scenario with a simplest mechanism
of neutrino--flavour oscillations. Below we show that this turns out
to be enough for the reduction of the theoretical values of the solar
neutrino fluxes in agreement with the experimental data and does not
demand a fine tuning of neutrino--flavour oscillation parameters.

The paper is organized as follows. In Section\,2 we give a cursory
outline of the Nambu--Jona--Lasinio model of light nuclei. In Sect.\,3
we calculate the reduced value of the solar core temperature and the
solar neutrino fluxes. The experimental data and theoretical results
are adduced in Tables 1--5. In the Conclusion we discuss the obtained
results.

\section{The Nambu--Jona--Lasinio model of light nuclei}
\setcounter{equation}{0}

\hspace{0.2in} Nowadays neither the PMA nor the EFT can provide a
dynamical enhancement of the astrophysical factor $S_{\rm pp}(0)$
relative to the recommended value $S^{\rm SSM}_{\rm pp}(0) =
4.00\times 10^{-25}\,{\rm MeV b}$ [4]. As has been pointed out in
Refs.\,[5--9] the contributions of low--energy nuclear forces are
taken into account with an accuracy better than 1$\%$\footnote{An
attempt to change the value of the astrophysical factor $S_{\rm
pp}(0)$ within the EFT has been recently undertaken by Kong and
Ravndal [23]. They have calculated the astrophysical factor $S^{\rm
KR}_{\rm pp}(0) = (4.31\pm 0.35)\times 10^{-25}\,{\rm MeV\,b}$ the
meanvalue of which is increased by about 8$\%$.}.

The required enhancement of the astrophysical factor $S_{\rm pp}(0)$
can be obtained within the Nambu--Jona--Lasinio model of light nuclei
[24--26], or differently the nuclear Nambu--Jona--Lasinio (NNJL)
model, invented for the description of low--energy nuclear forces at
the quantum field theoretic level. As has been shown in Ref.\,[24] the
NNJL model is fully motivated by QCD.  The deuteron appears in {\it
the nuclear phase of QCD} as a neutron--proton collective excitation,
a Cooper np--pair\,\footnote{In this connection we would like to refer
to the paper by Baldo, Lombardo and Schuck [27], where there has been
shown that the formation of the deuteron in heavy ion reactions at
intermediate energies goes through the superfluid Cooper pair phase to
a Bose deuteron gas.}, caused by a phenomenological local
four--nucleon interaction.  Strong low--energy interactions of the
deuteron coupled to itself and other particles are described in terms
of one--nucleon loop exchanges.  This allows to transfer nuclear
flavours from an initial to a final nuclear state by a minimal way and
to take into account contributions of nucleon--loop anomalies
determined completely by one--nucleon loop diagrams. The dominance of
contributions of nucleon--loop anomalies is justified in the large
$N_C$ approach to non--perturbative QCD with $SU(N_C)$ gauge group at
$N_C\to\infty$, where $N_C$ is the number of quark colours [24].

In Ref.\,[26] the NNJL model has been applied to the description of
low--energy nuclear forces for electromagnetic and weak nuclear
reactions with the deuteron. There have been calculated cross sections
and astrophysical factors for the neutron--proton radiative capture
(M1--capture) n + p $\to$ D + $\gamma$, the photomagnetic
disintegration of the deuteron $\gamma$ + D $\to$ n + p and weak
reactions of astrophysical interest such as: 1) the solar proton
burning p + p $\to$ D + e$^+$ + $\nu_{\rm e}$, 2) the pep--process p +
e$^-$ + p $\to$ D + $\nu_{\rm e}$ and 3) the reactions of neutrino and
anti--neutrino disintegration of the deuteron caused by charged
$\nu_{\rm e}$ + D $\to$ e$^-$ + p + p, $\bar{\nu}_{\rm e}$ + D $\to$
e$^+$ + n + n and neutral $\nu_{\rm e}(\bar{\nu}_{\rm e})$ + D $\to$
$\nu_{\rm e}(\bar{\nu}_{\rm e})$ + n + p weak currents.

In the NNJL model the deuteron couples to itself and other particles
through the nucleon axial--vector current $j^{\mu}(x)=
-i\,[\bar{p^c}(x)\gamma^{\mu} n(x) - \bar{n^c}(x)\gamma^{\mu} p(x)]$
with a coupling constant $g_{\rm V}$ and the nucleon tensor current
$J^{\mu\nu}(x)= \bar{p^c}(x)\sigma^{\mu\nu} n(x) -
\bar{n^c}(x)\sigma^{\mu\nu} p(x)$ with a coupling constant $g_{\rm T}$
connected with $g_{\rm V}$ by the relation $g_{\rm T} =
\sqrt{3/8}\,g_{\rm V}$ [24]. In turn, the coupling constant $g_{\rm
V}$ is determined by the electric quadrupole moment of the deuteron
$Q_{\rm D}$: $g^2_{\rm V} = 2\pi^2 Q_{\rm D} M^2_{\rm N}$ [24], where
$M_{\rm N} = 940\,{\rm MeV}$ is the nucleon mass.

The reaction of the neutron--proton radiative capture for thermal
neutrons n + p $\to$ D + $\gamma$ plays an important role for
nucleosynthesis in Early Universe [2]. For thermal neutrons the
reaction n + p $\to$ D + $\gamma$ is the M1--capture induced fully by
the magnetic M1--transition. The M1--capture n + p $\to$ D + $\gamma$
and the photomagnetic disintegration of the deuteron $\gamma$ + D
$\to$ n + p are related via time--reversal invariance. For the
evaluation of the amplitude of the M1--capture in the NNJL model [26]
we have taken into account both chiral one--meson loop corrections,
obtained in Chiral perturbation theory at the quark level (CHPT)$_q$
developed within the extended Nambu--Jona--Lasinio (ENJL) model with a
linear realization of chiral $U(3)\times U(3)$ symmetry [28], and the
$\Delta(1232)$ resonance treated as a Rarita--Schwinger field. We have
shown that within the experimental uncertainties of the definition of
the coupling constant of the ${\rm \pi \Delta N}$ interaction
off--mass shell of the $\Delta(1232)$ resonance described by the
parameter $Z$ the NNJL model fits well the experimental value of the
cross section for the neutron--proton radiative capture. At $Z = 1/2$
favoured theoretically [29] we have got the result $\sigma({\rm np \to
D\gamma})(T_{\rm n}) = 325.5\,{\rm m b}$, where $T_{\rm n} =
0.0253\,{\rm eV}$ is the kinetic energy of a thermal neutron in the
laboratory frame, agreeing with the experimental value $\sigma({\rm np
\to D\gamma})(T_{\rm n}) = (334.2\pm 0.5)\,{\rm m b}$ with an accuracy
about 3$\%$ [26]. Hence, we argue that in the NNJL model as well as in
the EFT, developing a quantum field theoretic approach to the
description of low--energy nuclear forces but, unlike the NNJL model,
in relativistically non--covariant way, all corrections and
contributions to the amplitudes of low--energy nuclear reactions are
under the control. We would like to note that due to the loss of
relativistic covariance nucleon--loop anomalies are ill--defined in
the EFT that makes impossible any application of a mechanism of
nucleon--loop anomalies to the correct description of low--energy
nuclear forces. In turn, in the NNJL model, the relativistically
covariant quantum field theoretic model, nucleon--loop anomalies are
well defined and give dominant contributions.

As has been shown in Ref.[26] the contribution of the nucleon tensor
current enters to the cross sections for weak nuclear reactions with
the deuteron with an arbitrary parameter $\bar{\xi}$\,\footnote{The
appearance of an arbitrary parameter in the astrophysical factor for
the solar proton burning the value of which is governed by the
experimental data on the cross sections for the low--energy reactions
of neutrino and anti--neutrino disintegration of the deuteron has been
argued recently by Kong and Ravndal [23] in the EFT approach.}. Due to
isotopical invariance of low--energy nuclear forces this parameter is
the same for all low--energy weak nuclear reactions involving pp, nn
or np states.

At zero contribution of the nucleon tensor current, $\bar{\xi} =
0$\,\footnote{By checking the formulas of the cross sections for weak
nuclear reactions with the deuteron calculated in Ref.[26] one can see
that the value $\bar{\xi} = -2$ leads to the same result as well as
the value $\bar{\xi} = 0$. Below we would call $\bar{\xi}^{\,\rm PT}$
and $\bar{\xi}^{\,\rm NPT}$ as a perturbative, close to $\bar{\xi} =
0$, and a non--perturbative, close to $\bar{\xi} = -2$, solution for
the parameter $\bar{\xi}$, respectively.}, we have obtained [26]: 1)
the astrophysical factor for the solar proton burning $S_{\rm pp}(0) =
4.08\times 10^{-25}\,{\rm MeV\, b}$ agreeing well with the recommended
one [4]\,\footnote{It is interesting too to compare the value of
$S_{\rm pp}(0)$ calculated in the NNJL model caused by the
contribution of the nucleon axial--vector current with the
old--fashioned result for $S_{\rm pp}(0)$ obtained by Bahcall and
Ulrich in 1988 [30]: $S^{\rm BU}_{\rm pp}(0) = 4.07\times
10^{-25}\,{\rm MeV\,b}$.}, 2) the astrophysical factor for the
pep--process $S_{\rm pep}(0)$ relative to $S_{\rm pp}(0)$ in complete
agreement with the result obtained by Bahcall and May
[31]\,\footnote{We would like to emphasize that the ratio $S_{\rm
pep}(0)/S_{\rm pp}(0)$ does not depend on the parameter
$\bar{\xi}$. Therefore, the result obtained in Ref.[26] for the ratio
$S_{\rm pep}(0)/S_{\rm pp}(0)$ is valid for arbitrary contribution of
the nucleon tensor current.}, 3) the cross section for the neutrino
disintegration of the deuteron $\nu_{\rm e}$ + D $\to$ e$^-$ + p + p
caused by charged weak current and 4) the cross sections for the
anti--neutrino disintegration of the deuteron $\bar{\nu}_{\rm e}$ + D
$\to$ e$^+$ + n + n and $\bar{\nu}_{\rm e}$ + D $\to$ $\bar{\nu}_{\rm
e}$ + n + p caused by charged and neutral weak currents, respectively,
and averaged over anti--neutrino spectrum in a reasonable agreement
with recent experimental data obtained by the Reines's experimental
group [32].

Hence, in the NNJL model at $\bar{\xi} = 0$ (or $\bar{\xi} = - 2$ that
is the same) we get a dynamics of low--energy nuclear forces for the
description of weak nuclear reactions with the deuteron in agreement
with the recommendations of the SSM.

Of course, one can develop a scenario of the description of
low--energy nuclear forces contributing to weak nuclear reactions with
the deuteron when the parameter $\bar{\xi} \not= 0$ (or that is the
same $\bar{\xi} \not= - 2$) and tune this parameter in such a way in
order to get an enhancement of the value of the astrophysical factor
$S_{\rm pp}(0)$ relative to the recommended one $S^{\rm SSM}_{\rm
pp}(0) = 4.00\times 10^{-25}\,{\rm MeV\,b}$ [4].

Thus, the NNJL model [24--26]\,\footnote{As well as the result
obtained by Kong and Ravndal [23] in the EFT approach.} gives a hint
that there is a dynamical reason, a non--trivial contribution of
low--energy nuclear forces induced by the nucleon tensor current
coupled to the deuteron, for an enhancement of the astrophysical
factor for the solar proton burning leading, in turn, to a reduction
of the solar core temperature.

The main goal of this paper is to analyse the consistency of this
reduction of the solar core temperature with the SSM,
helioseismological data, the experimental data on the solar neutrino
fluxes and experimental data on cross sections for the reactions of
the anti--neutrino disintegration of the deuteron [32].  On this way
in order to be close as much as it is possible to the SSM we would
like to accentuate that the theoretical values of the solar neutrino
fluxes, their dependence on the solar core temperature and the
relationship between the changes of the astrophysical factor for the
solar proton burning and the solar core temperature would be taken
from the SSM [11]. Thereby, the theoretical accuracy of our results
for solar neutrino fluxes should coincide with the theoretical
accuracy of the SSM of the calculation of the solar neutrino fluxes.
Henceforth, only in order to distinguish all theoretical quantities
obtained by virtue of the reduction of the solar core temperature from
those calculated in the SSM we suggest to use the label ASM that
stands for the abbreviation for the Alternative Solar Model. The ASM
leaves unchanged all results of the SSM save the value of the solar
core temperature reduced in the ASM with respect to that recommended
by the SSM [4].

\section{Solar neutrino fluxes}
\setcounter{equation}{0}

\hspace{0.2in} As has been shown in Ref.\,[26] a non--trivial
contribution of the nucleon tensor current [24] caused by the
interaction
\begin{eqnarray}\label{label3.1}
\delta {\cal L}_{\rm npD}(x) = \sqrt{\frac{3}{8}}\,\frac{g_{\rm
V}}{2M_{\rm N}}\,[\bar{p^c}(x)\sigma^{\mu\nu}n(x) -
\bar{n^c}(x)\sigma^{\mu\nu}p(x)]\,D^{\dagger}_{\mu\nu}(x) + {\rm h.c.}
\end{eqnarray}
changes\,\footnote{Here $D^{\dagger}_{\mu\nu}(x) =
\partial_{\mu}D^{\dagger}_{\nu}(x) -
\partial_{\nu}D^{\dagger}_{\mu}(x)$ and $D^{\dagger}_{\nu}(x)$ is the
operator of the interpolating field of the deuteron, $\bar{p^c}(x) =
p^T(x)C$ and $\bar{n^c}(x) = n^T(x)C$ with $C$ is a charge conjugation
matrix and $T$ is a transposition.} the astrophysical factor 
$S_{\rm pp}(0)$ for the
solar proton burning p + p $\to$ D + e$^+$ + $\nu_{\rm e}$ as follows
\begin{eqnarray}\label{label3.2}
S^{\rm ASM}_{\rm pp}(0) =  (1 + \bar{\xi}\,)^2\times 4.08\times 10^{-25}\,{\rm
MeV\,b},
\end{eqnarray}
where $\bar{\xi}$ is an arbitrary parameter (see Appendix of
Ref.\,[26])\,\footnote{The same factor $(1 + \bar{\xi})^2$ appears in
the astrophysical factor for the pep--process p + e$^-$ + p $\to$ D +
$\nu_{\rm e}$ and, due to the isotopical invariance of low--energy
nuclear forces, in the cross sections for neutrino and anti--neutrino
disintegration of the deuteron $\nu_{\rm e}$ + D $\to$ e$^-$ + p + p,
$\bar{\nu}_{\rm e}$ + D $\to$ e$^+$ + n + n and $\bar{\nu}_{\rm
e}(\nu_{\rm e})$ + D $\to$ $\bar{\nu}_{\rm e}(\nu_{\rm e})$ + n + p
[26].}.

Below we develop a scenario of a dynamics of low--energy nuclear
forces providing $(1 + \bar{\xi})^2 \ge 1$ and analyse a consistency
of this dynamics with the helioseismological data [10], the
experimental data on the solar neutrino fluxes [12--16] and the cross
sections for the reactor anti--neutrino disintegration of the deuteron
[32]

The changes $\Delta T_c$ of the solar core temperature $T_c$ and
$\Delta S_{\rm pp}(0)$ of the astrophysical factor $S_{\rm pp}(0)$ for
the solar proton burning with respect to the values $T^{\rm SSM}_c =
1.574\times 10^7\,{\rm K}$ [9] and $S^{\rm SSM}_{\rm pp}(0) =
4.00\times 10^{-25}\,{\rm MeV\,b}$ recommended by the SSM are related
by [9]
\begin{eqnarray}\label{label3.3}
\frac{\Delta T_c}{T^{\rm SSM}_c} = -\,0.15\,\frac{\Delta S_{\rm
pp}(0)}{S^{\rm SSM}_{\rm pp}(0)}.
\end{eqnarray}
According to the helioseismological data [10] the maximum value of the
astrophysical factor can be equal to $S^{\rm max}_{\rm pp}(0) =
1.18\times S^{\rm SSM}_{\rm pp}(0) = 1.18\times 4.00\times
10^{-25}\,{\rm MeV\,b} = 4.72\times 10^{-25}\,{\rm MeV\,b}$. From
Eq.(\ref{label3.3}) it is seen that the maximum value of the
astrophysical factor defines the minimum value of the solar core
temperature. The helioseismological data give also the lower bound on
the astrophysical factor [10]: $S^{\rm min}_{\rm pp}(0) = 0.94\times
S^{\rm SSM}_{\rm pp}(0) = 0.94\times 4.00\times 10^{-25}\,{\rm MeV\,b}
= 3.76\times 10^{-25}\,{\rm MeV\,b}$. The minimum value of the
astrophysical factor corresponds to the maximal solar core temperature
which is greater than that predicted by the SSM, $T^{\rm SSM}_c =
1.574\times 10^7\,$K.  Since for the temperatures higher than $T^{\rm
SSM}_c$ solar neutrino fluxes become increased with respect to those
predicted by the SSM, so that we would consider only temperatures
lower than $T^{\rm SSM}_c$ and astrophysical factors greater than the
recommended one $S^{\rm SSM} = 4.00\times 10^{-25}\,{\rm MeV\,b}$.

In the NNJL model due to the isotopical invariance of low--energy
nuclear forces the contribution of the nucleon tensor current
enters into the cross sections for the neutrino and anti--neutrino
disintegration of the deuteron with the same parameter $\bar{\xi}$
[26]. This means that the enhancement of the astrophysical factor
$S_{\rm pp}(0)$ in the NNJL model should be restricted not only by the
helioseismological data but also the experimental data on the cross
sections for the neutrino and anti--neutrino disintegration of the
deuteron at low energies [26].

First, let us find out the maximal reduction of the solar neutrino
fluxes caused by the cooling of the solar core up to the minimal
temperature. By equating the astrophysical factor calculated in the
NNJL model to the maximum value $S^{\rm ASM}_{\rm pp}(0) = S^{\rm max}_{\rm
pp}(0) = 4.72\times 10^{-25}\,{\rm MeV\,b}$ allowed by the
helioseismic data [10], we obtain the minimal value of the solar core
temperature
\begin{eqnarray}\label{label3.4}
T^{\rm ASM}_c = 1.533\times 10^7\,{\rm K}.
\end{eqnarray}
Thus, due to the helioseismological constraint the minimal value of the
solar core temperature cannot be less than the recommended one by
about 2.7$\%$\,\footnote{For the parameter $\bar{\xi}$ we get
perturbative $\bar{\xi}^{\,\rm PT} = 0.077$ and
non--perturbative $\bar{\xi}^{\,\rm NPT} = -\,2.077$
solutions. The results of the SSM can be restored at $\bar{\xi}^{\,\rm
PT} = 0$ and $\bar{\xi}^{\,\rm NPT} = -\,2$,
respectively.}.

In Tables 1 and 2 we adduce the experimental data on the solar
neutrino fluxes and the theoretical values of the solar neutrino
fluxes predicted by the SSM [3,11] and normalized to the recommended
value of the astrophysical factor [4]. The theoretical accuracies of
the solar neutrino fluxes predicted by the SSM (see Table 2) are equal
to $(+15.6\%,-13.0\%)$, $(+ 6.2\%,-4.7\%)$ and $(+19.4\%,-13.6\%)$ for
the HOMESTAKE, GALLEX, SAGE and SUPERKAMIOKANDE experiments,
respectively. This implies that the theoretical solar neutrino fluxes
reduced by virtue of the reduction of the solar core temperature
should be defined with the same accuracy. Thereby, the description of
the experimental data within the ASM inheriting the accuracy of the
SSM cannot be carried out with an accuracy better than
$(+15.6\%,-13.0\%)$, $(+ 6.2\%,-4.7\%)$ and $(+19.4\%,-13.6\%)$ for
the data by HOMESTAKE, GALLEX, SAGE and SUPERKAMIOKANDE
Collaborations, respectively.

By using a temperature dependence of the solar neutrino fluxes
obtained by Bahcall and Ulmer [33]: $\Phi({\rm pp}) \propto
T^{-1.1}_c$, $\Phi({\rm pep}) \propto T^{-2.4}_c$, $\Phi({^7}{\rm Be})
\propto T^{\,10}_c$, $\Phi({^8}{\rm B}) \propto T^{\,24}_c$,
$\Phi({^{13}}{\rm N}) \propto T^{\,24.4}_c$ and $\Phi({^{15}}{\rm O})
\propto T^{\,27.1}_c$ we can calculate the solar neutrino fluxes for
the reduced solar core temperature Eq.(\ref{label3.4}).  The new
values of the solar neutrino fluxes we have adduced in Table 3.

It is seen that the solar neutrino fluxes calculated for the solar
core temperature $T^{\rm ASM}_c = 1.533\times 10^7\,$K are still not
enough decreased in order to satisfy the experimental data. The next
step for the reduction of the solar neutrino fluxes taking place
outside the solar core is in the attraction of neutrino--flavour
oscillations [17--20]. We would follow a simplest scenario of vacuum
two--flavour neutrino oscillations [17,18]. By virtue of the vacuum
two--flavour neutrino oscillations $\nu_{\rm e} \leftrightarrow
\nu_{\mu}$ the theoretical solar neutrino fluxes should be multiplied
by a factor [17--20]
\begin{eqnarray}\label{label3.5}
P_{\nu_{\rm e} \to  \nu_{\rm e}}(E_{\nu_{\rm e}}) = 1 - \frac{1}{2}\,\sin^2
2\,\theta\,\Bigg(1 - \cos\frac{\Delta m^2 L}{2E_{\nu_{\rm e}}}\Bigg),
\end{eqnarray}
where $\Delta m^2 = m^2_{\nu_{\mu}} - m^2_{\nu_{\rm e}}$, $L$ is the
distance of the neutrino's travel, $E_{\nu_{\rm e}}$ is a neutrino
energy and $\theta$ is a neutrino--flavour mixing angle [17]. After
the averaging over energies and keeping $L$ of order of the Sun--Earth
distance the theoretical solar neutrino fluxes calculated for the
reduced solar core temperature Eq.(\ref{label3.4}) become multiplied
by a factor [18]
\begin{eqnarray}\label{label3.6}
\overline{P_{\nu_{\rm e} \to \nu_{\rm e}}(E_{\nu_{\rm e}})} = 1 -
\frac{1}{2}\,\sin^2 2\,\theta.
\end{eqnarray}
The result of the integration over energies Eq.(\ref{label3.6}) can be
valid only if the quantity $\Delta m^2 L/2 E_{\nu_{\rm e}}$ obeys the
constraint
\begin{eqnarray}\label{label3.7}
\frac{\Delta m^2 L}{2 E_{\nu_{\rm e}}} \gg 1.
\end{eqnarray}
If we would like to get the factor Eq.(\ref{label3.6}) for all solar
neutrino fluxes including the ${^8}{\rm B}$ neutrinos, the upper bound
on the neutrino energies should coincide with the upper bound on the
${^8}{\rm B}$ neutrino energy spectrum equal to $E_{\nu_{\rm e}} =
15\,{\rm MeV}$ [1]. As the Sun--Earth distance $L$ amounts to $L =
1.496\times 10^{13}\,{\rm cm} = 7.581\times 10^{23}\,{\rm MeV^{-1}}$,
the inequality Eq.(\ref{label3.7}) gives the lower bound on $\Delta
m^2$:
\begin{eqnarray}\label{label3.8}
\Delta m^2 \gg  4 \times 10^{-11}\,{\rm eV}^2.
\end{eqnarray}
Hence, in order to get a correct agreement with the experimental data
on the solar neutrino fluxes measured by HOMESTAKE, GALLEX and SAGE we
do not need to make {\it a fine tuning of the neutrino--flavour
oscillation parameter $\Delta m^2$} [34] and to have much more
information about $\Delta m^2$ save that given by Eq.(\ref{label3.8}).

The value of the mixing angle $\sin^2 2\theta$ we can get fitting, for
example, the meanvalue of the neutrino flux measured by HOMESTAKE
Collaboration. This gives $\sin^2 2\theta = 0.838$.  The solar
neutrino fluxes reduced by virtue of vacuum two--flavour neutrino
oscillations are adduced in Table 4. One can see that the theoretical
solar neutrino fluxes fit well the experimental data by GALLEX and
SAGE Collaborations.

The theoretical value of the solar neutrino flux $\Phi^{\rm ASM}_{\rm
SK}({^8}{\rm B})$ fitting the experimental data of SUPERKAMIOKANDE
Collaboration is related to the solar ${^8}{\rm B}$ neutrino flux
$\Phi^{\rm ASM}({^8}{\rm B})$. This relation can be derived by
following Bahcall {\it et al.} [35]. With an accuracy better than 2$\%$
the theoretical expression for the solar neutrino flux $\Phi^{\rm ASM}_{\rm
SK}({^8}{\rm B})$ is given by
\begin{eqnarray}\label{label3.9}
\Phi^{\rm ASM}_{\rm SK}({^8}{\rm B}) = \Bigg(1 - \,
\frac{4\,\sin^22\theta\,\sin^2\theta_{\rm W}}{(1 \,+\, 2\,\sin^2\theta_{\rm
W})^2}\Bigg)\,\Phi^{\rm ASM}({^8}{\rm B}) = (1.73^{+0.34}_{-0.24})\times
10^6\,{\rm cm^{-2}s^{-1}},
\end{eqnarray}
where $\theta_{\rm W}$ is the Weinberg's mixing angle of the Standard
electroweak model equal to $\sin^2\theta_{\rm W} = 0.225$ [16]. The
theoretical value $\Phi^{\rm ASM}_{\rm SK}({^8}{\rm B}) =
(1.73^{+0.34}_{-0.24})\times 10^6\,{\rm cm^{-2}s^{-1}}$ is comparable
with the experimental one $\Phi^{\exp}_{\rm SK}({^8}{\rm B}) =
(2.40\pm 0.09)\times 10^6\,{\rm cm^{-2}s^{-1}}$ [16] with an accuracy
about 30$\%$. In order to get the theoretical solar neutrino fluxes
fitting the experimental data with accuracies $(+15.6\%, -13.0\%)$,
$(+6.2\%, -4.7\%)$ and $(+19.4\%, -13.6\%)$ for HOMESTAKE, GALLEX
(SAGE) and SUPERKAMIOKANDE Collaborations, respectively, it is
sufficient to diminish the value of the mixing angle up to
$\sin^22\theta = 0.607$. This yields: $S^{\rm ASM}_{\rm Cl} =
(3.07^{+0.49}_{-0.40})\,{\rm SNU}$, $S^{\rm ASM}_{\rm Ga} =
(77.98^{+4.83}_{-3.67})\,{\rm SNU}$ and $\Phi^{\rm ASM}_{\rm
SK}({^8}{\rm B}) = (2.00^{+0.39}_{-0.27})\times 10^6\,{\rm
cm^{-2}s^{-1}}$. The errours of the theoretical values of the solar
neutrino fluxes coincide with the theoretical errours of the SSM
inherited by the ASM.

Since the theoretical solar neutrino fluxes fit reasonably well the
experimental data, so that this evidences the solution of the SNP's in
the form formulated by Bahcall [11]. 

We would like to emphasize that the agreement between the theoretical
solar neutrino fluxes and the experimental ones has been reached
without a fine tuning of the neutrino--flavour oscillation parameters
$\Delta m^2$ and $\sin^22\theta$.

Now we have to analyse how the maximal enhancement of the
astrophysical factor allowed by the helioseismological data and
providing the solution of the SNP's affects the theoretical values of
the cross sections for the reactor anti--neutrino disintegration of
the deuteron [26,32].  On this way one can find that the enhancement
of the astrophysical factor up to the value $S^{\rm ASM}_{\rm pp}(0) =
4.72\times 10^{-25}\,{\rm MeV\,b}$ leads to the theoretical values of
the cross sections for the reactions $\bar{\nu}_{\rm e}$ + D $\to$
e$^+$ + n + n and $\bar{\nu}_{\rm e}$ + D $\to$ $\bar{\nu}_{\rm e}$ +
n + p fitting the meanvalues of the experimentally measured cross
sections with an accuracy 1.75$\,\sigma$ and 1.55$\,\sigma$,
respectively [26,32].  Such an agreement being reasonable in principle
can be improved by diminishing the value of the astrophysical factor
$S^{\rm ASM}_{\rm pp}(0) = 4.72\times 10^{-25}\,{\rm MeV\,b}$.

One of the ways for the optimization of the enhancement of the
astrophysical factor can be, for example, the following. Let the
theoretical solar neutrino fluxes be functions of two variables
$(\sin^22\theta, \lambda)$, where $\lambda$ is related to the
theoretical solar neutrino flux $\Phi^{\rm ASM}_{\rm SK}({^8}{\rm B})
= \lambda\times 10^6\,{\rm cm^{-2}s^{-1}}$ fitting the experimental
data by SUPERKAMIOKANDE Collaboration. From Eq.(\ref{label3.9}) the
theoretical solar ${^8}{\rm B}$ neutrino flux $\Phi^{\rm ASM}({^8}{\rm
B})$ can be expressed in terms of $\Phi^{\rm ASM}_{\rm SK}({^8}{\rm
B}) = \lambda\times 10^6\,{\rm cm^{-2}s^{-1}}$ and $\sin^22\theta$.
At $\sin^2\theta_{\rm W} = 0.225$ we obtain
\begin{eqnarray}\label{label3.10}
\Phi^{\rm ASM}({^8}{\rm B})= \lambda\times 10^6\,(1-
0.428\,\sin^22\theta)^{-1}\,{\rm cm^{-2}s^{-1}}.
\end{eqnarray}
The ratio of the solar core temperatures $T^{\rm ASM}_c/T^{\rm SSM}_c$
is then defined by
\begin{eqnarray}\label{label3.11}
\frac{T^{\rm ASM}_c}{T^{\rm SSM}_c} = 0.934\,\lambda^{1/24}(1 - 0.428\,\sin^22\theta)^{-1/24}.
\end{eqnarray}
The solar neutrino fluxes as functions of $\sin^22\theta$ and
$\lambda$ measured in $10^{10}\,{\rm cm^{-2}s^{-1}}$ read [11]:
\begin{eqnarray}\label{label3.12}
\Phi^{\rm ASM}({\rm pp})&=&6.403\,\lambda^{-1.1/24}(1- 0.428\,
\sin^22\theta)^{1.1/24},\nonumber\\ \Phi^{\rm ASM}({\rm
pep})&=&1.638\times 10^{-2}\,\lambda^{-1/10}(1-
0.428\,\sin^22\theta)^{1/10},\nonumber\\ \Phi^{\rm ASM}({^7}{\rm
Be})&=&2.425\times 10^{-1}\,\lambda^{10/24}(1-
0.428\,\sin^22\theta)^{- 10/24},\nonumber\\ \Phi^{\rm ASM}({^8}{\rm
B})&=&1.000\times 10^{-4}\,\lambda\,(1-
0.428\,\sin^22\theta)^{-1},\nonumber\\ \Phi^{\rm ASM}({^{13}}{\rm
N})&=&1.144\times 10^{-2}\,\lambda^{24.4/24}(1-
0.428\,\sin^22\theta)^{- 24.4/24},\nonumber\\ \Phi^{\rm
ASM}({^{15}}{\rm O})&=&0.836\times 10^{-2}\,\lambda^{27.1/24}(1-
0.428\,\sin^22\theta)^{- 27.1/24}.
\end{eqnarray}
The theoretical expressions for the solar neutrino fluxes measured by
HOMESTAKE, GALLEX and SAGE experiments are given by [11]
\begin{eqnarray}\label{label3.13}
&&S^{\rm ASM}_{\rm Cl} = (1 - 0.5\,\sin^22\theta)\nonumber\\
&&\times\,[0.236\,\lambda^{-1/10}(1-
0.428\,\sin^22\theta)^{1/10} + 0.582\,\lambda^{10/24}(1-
0.428\,\sin^22\theta)^{- 10/24} \nonumber\\
&&+ 0.019\,\lambda^{24.4/24}(1-
0.428\,\sin^22\theta)^{- 24.4/24} + 0.063\,\lambda^{27.1/24}(1-
0.428\,\sin^22\theta)^{- 27.1/24}\nonumber\\
&&+ 1.146\,\lambda\,(1-
0.428\,\sin^22\theta)^{-1}],\nonumber\\
&&S^{\rm ASM}_{\rm Ga} =(1 - 0.5\,\sin^22\theta)\nonumber\\
&&\times\,[75.043\,\lambda^{-1.1/24}(1-
0.428\,\sin^22\theta)^{1.1/24} + 3.300\,\lambda^{-1/10}(1-
0.428\,\sin^22\theta)^{1/10}\nonumber\\
&&+ 17.380\,\lambda^{10/24}(1-
0.428\,\sin^22\theta)^{- 10/24} + 0.700\,\lambda^{24.4/24}(1-
0.428\,\sin^22\theta)^{- 24.4/24}\nonumber\\
&&+ 0.943\,\lambda^{27.1/24}(1-
0.428\,\sin^22\theta)^{- 27.1/24} + 2.408\,\lambda\,(1-
0.428\,\sin^22\theta)^{-1}],
\end{eqnarray}
where the factor $(1 - 0.5\,\sin^22\theta)$ takes into account the
contribution of vacuum two--flavour neutrino oscillations.

For the fit of experimental data on the solar neutrino fluxes by the
theoretical expressions Eq.(\ref{label3.13}) we can feel ourselves to
be constrained only by the requirement of the description of the
experimental data by GALLEX and SAGE Collaborations with an accuracy not
worse than $(+ 6.2\%, - 4.7\%)$ inherited from the SSM, as the solar
neutrino fluxes for HOMESTAKE and SUPERKAMIOKANDE Collaborations are
determined in the SSM with a much worse accuracy.

The theoretical solar neutrino fluxes fitting the experimental data by
GALLEX and SAGE Collaborations with an accuracy $(+ 6.2\%, - 4.7\%)$ can
be obtained at $\lambda = 2.12$ and $\sin^22\theta = 0.780$:
$\Phi^{\rm ASM}_{\rm SK}({^8}{\rm B}) = (2.12^{+0.41}_{-0.29})\times
10^6\,{\rm cm^{-2}s^{-1}}$, $S^{\rm ASM}_{\rm Cl} =
(3.11^{+0.49}_{-0.40})\,{\rm SNU}$ and $S^{\rm ASM}_{\rm Ga} =
(70.56^{+4.38}_{-3.32})\,{\rm SNU}$. The errours are the theoretical
uncertainties of the calculation of the solar neutrino fluxes in the
SSM [11]. The solar neutrino fluxes given by Eqs.(\ref{label3.12}) and
(\ref{label3.13}) and calculated at $\lambda = 2.12$ and $\sin^22\theta
= 0.780$ are adduced in Table 5.

If there would be allowed to fit the experimental data by GALLEX and
SAGE Collaborations with an accuracy worse than $(+ 6.2\%, - 4.7\%)$,
the region of variables $(\sin^22\theta, \lambda)$ would become much
broader. For example, at the maximum mixing angle $\sin^22\theta = 1$
and $\lambda = 2.12$ we obtain: $\Phi^{\rm ASM}_{\rm SK}({^8}{\rm B})
= (2.12^{+0.41}_{-0.29})\times 10^6\,{\rm cm^{-2}s^{-1}}$, $S^{\rm
ASM}_{\rm Cl} = (2.90^{+0.45}_{-0.38})\,{\rm SNU}$ and $S^{\rm
ASM}_{\rm Ga} = (59.64^{+3.70}_{-2.80})\, {\rm SNU}$. Our prediction
for the low--energy solar neutrino flux agrees with the experimental
data by GALLEX and SAGE Collaborations with an accuracy about 20$\%$
(see Table 1) and fits well the experimental data by GNO
Collaboration. The experimental value of the high--energy
solar neutrino flux measured by HOMESTAKE Collaboration is fitted with
an accuracy about 12$\%$.

If there would be set $\sin^22\theta = 1$ and $\lambda = 2.40$ that
corresponds to the meanvalue of the experimental flux measured by
SUPERKAMIOKANDE Collaboration, there would have been obtained the
following theoretical predictions: $\Phi^{\rm ASM}_{\rm SK}({^8}{\rm
B}) = (2.40^{+0.47}_{-0.33})\times 10^6\,{\rm cm^{-2}s^{-1}}$, $S^{\rm
ASM}_{\rm Cl} = (3.24^{+0.51}_{-0.42})\,{\rm SNU}$ and $S^{\rm
ASM}_{\rm Ga} = (61.30^{+3.80}_{-2.90})\,{\rm SNU}$ fitting the
experimental data on the high--energy (HOMESTAKE) and low--energy
(GALLEX and SAGE) solar neutrino fluxes with an accuracy about 25$\%$. In
turn, at $\sin^22\theta = 0.780$ and $\lambda = 2.40$ one obtains:
$\Phi^{\rm ASM}_{\rm SK}({^8}{\rm B}) = (2.40^{+0.47}_{-0.33})\times
10^6\,{\rm cm^{-2}s^{-1}}$, $S^{\rm ASM}_{\rm Cl} =
(3.46^{+0.54}_{-0.45})\,{\rm SNU}$ and $S^{\rm ASM}_{\rm Ga} =
(72.33^{+4.49}_{-3.04})\,{\rm SNU}$.

This testifies a consistency of the experimental solar neutrino fluxes
with theoretical fluxes given by Eqs.(\ref{label3.12}) and
(\ref{label3.13}) calculated in the SSM with a reduced solar core
temperature and supplemented by the scenario of vacuum two--flavour
solar neutrino oscillations.

The reduced solar core temperature $T^{\rm ASM}_c$, the astrophysical
factor for the solar proton burning $S^{\rm ASM}_{\rm pp}(0)$ and the
enhancement factor $S^{\rm ASM}_{\rm pp}(0)/S^{\rm SSM}_{\rm pp}(0)$
calculated at $\lambda = 2.12$ and $\sin^22\theta = 0.780$ are equal
to
\begin{eqnarray}\label{label3.14}
T^{\rm ASM}_c &=& (1.549^{+0.005}_{-0.016})\times 10^7\,{\rm
K},\nonumber\\ S^{\rm ASM}_{\rm pp}(0) &=&
(4.42^{+0.30}_{-0.08})\times 10^{-25}\,{\rm MeV\,b},\nonumber\\
\frac{\textstyle S^{\rm ASM}_{\rm pp}(0)}
{\textstyle S^{\rm SSM}_{\rm pp}(0)}&=&1.11^{+0.07}_{-0.02},
\end{eqnarray}
where the errours are defined by the theoretical uncertainties of the
solar ${^8}{\rm B}$ neutrino flux calculated in the SSM\,\,\footnote{For
the parameter $\bar{\xi}$ we obtain $\bar{\xi}^{\,\rm PT} =
0.041^{+0.031}_{-0.009}$ and $\bar{\xi}^{\,\rm NPT} =
-\,2.041^{+0.031}_{-0.009}$.}.

The theoretical values of the cross sections for the anti--neutrino
disintegration of the deuteron related to the astrophysical factors
corresponding to the theoretical solar neutrino fluxes calculated at
$(\sin^22\theta = 0.780, \lambda = 2.12)$, $(\sin^22\theta = 1,
\lambda = 2.12)$, $(\sin^22\theta = 1, \lambda = 2.40)$ and
$(\sin^22\theta = 0.780, \lambda = 2.40)$ fit the meanvalues of the
experimentally measured cross sections for the reaction
$\bar{\nu}_{\rm e}$ + D $\to$ e$^+$ + n + n and $\bar{\nu}_{\rm e}$ +
D $\to$ $\bar{\nu}_{\rm e}$ + n + p with an accuracy better than
1.32$\,\sigma$ and 1$\,\sigma$, respectively [26,32].

\section{Conclusion}
\setcounter{equation}{0}

\hspace{0.2in} We have shown that the scenario of a dynamics of
low--energy nuclear forces leading to the reduction of the solar core
temperature provides a reasonable theoretical foundation for the
solution of the SNP's in the form formulated by Bahcall [11]. Really,
the SSM with the reduced solar core temperature and supplemented by
the scenario of vacuum two--flavour neutrino oscillations $\nu_{\rm e}
\leftrightarrow \nu_{\mu}$ during the travel of solar neutrinos to the
Earth proposed by Gribov and Pontecorvo [17] admits a possibility to
calculate the theoretical solar neutrino fluxes fitting experimental
data without a fine tuning of neutrino--flavour oscillation parameters
$\Delta m^2$ and the mixing angle $\sin^22\theta$. As has been shown
due to the reduction of the solar neutrino fluxes in the solar core
for the simultaneous description of the experimental data on the solar
neutrino fluxes with an accuracy not worse than the theoretical
accuracy of the SSM it is sufficient to know only that $\Delta m^2 \gg
4\times 10^{-11}\,{\rm eV}^2$ and $\sin^22\theta \ge 0.780$.  The
former makes the application of neutrino--flavour oscillations to be
much more flexible with respect to different experimental constraints
on the parameters of the neutrino--flavour oscillations [36].

The theoretical solar neutrino fluxes given by Eqs.(\ref{label3.12})
and (\ref{label3.13}) and calculated at $\Delta m^2 \gg 4\times
10^{-11}\,{\rm eV}^2$ and $\sin^22\theta = 0.780$ read: $S^{\rm
ASM}_{\rm Cl} = (3.11^{+0.49}_{-0.40})\,{\rm SNU}$, $S^{\rm ASM}_{\rm
Ga} = (70.56^{+4.38}_{-3.32})\,{\rm SNU}$ and $\Phi^{\rm AS M}_{\rm
SK}({^8}{\rm B}) = (2.12^{+0.41}_{-0.29})\times 10^6\,{\rm
cm^{-2}s^{-1}}$.  For the ratios of Experiment\,:\,Theory we obtain
the numbers
\begin{eqnarray}\label{label4.1}
\frac{\textstyle S^{\exp}_{\rm Cl}}{\textstyle S^{\rm ASM}_{\rm
Cl}}&=&0.82^{+0.15}_{-0.13},\nonumber\\ \frac{\textstyle S^{\exp}_{\rm
Ga}}{\textstyle S^{\rm ASM}_{\rm
Ga}}&=&1.05^{+0.12}_{-0.11},\nonumber\\ \frac{\textstyle
\Phi^{\exp}_{\rm SK}({^8}{\rm B})}{\textstyle \Phi^{\rm ASM}_{\rm
SK}({^8}{\rm B})}&=&1.13^{+0.22}_{-0.14}
\end{eqnarray}
that are comparable with unity.

This reconciles the experimental data of the solar neutrino fluxes by
HOMESTAKE, GALLEX, SAGE and SUPERKAMIOKANDE Collaborations with the
theoretical predictions and solves the SNP's in the from formulated by
Bahcall [11].

The agreement with the experimental data by KAMIOKANDE Collaboration
$\Phi^{\rm K}_{\exp} =(2.80\pm 0.36)\times 10^6\,{\rm cm^{-2}s^{-1}}$
can be obtained within an accuracy about 15$\%$, where the error is
defined by $\sigma = \sqrt{({\rm stat.})^2 +({\rm syst.})^2}$.

The theoretical solar ${^8}{\rm B}$ and ${^7}{\rm Be}$ neutrino fluxes
defined by Eq.(\ref{label3.12}) and calculated at $\sin^2 2\theta =
0.780$ for the HOMESTAKE experiments are equal to $\Phi^{\rm ASM}_{\rm
Cl}({^8}{\rm B}) = (2.22^{+0.43}_{-0.30})\,{\rm SNU}$ and $\Phi^{\rm ASM}_{\rm
Cl}({^7}{\rm Be}) = (0.57^{+0.11}_{-0.08})\,{\rm SNU}$ and agree
reasonably well with recent data by HOMESTAKE [14] (Lande, Neutrino
2000): $\Phi^{\exp}_{\rm Cl}({^8}{\rm B}) = 2.16\,{\rm SNU}$ and
$\Phi^{\exp}_{\rm Cl}({^7}{\rm Be}) = 0.4\,{\rm SNU}$.

Thus, we have shown that the experimental data on the solar neutrino
fluxes measured by HOMESTAKE, GALLEX, SAGE, KAMIOKANDE and
SUPERKAMIOKANDE Collaborations are consistent with both each other and
the theoretical solar neutrino fluxes calculated in the SSM for the
reduced solar core temperature and supplemented by a scenario of
vacuum two--flavour neutrino oscillations.

The enhancement of the astrophysical factor being necessary for the
reduction of the solar neutrino fluxes in the solar core is caused by
the contribution of the nucleon tensor current coupled to the deuteron
and depends on the parameter $\bar{\xi}$ [26]. As has been shown in
Ref.[26] this parameter enters to the cross sections for the neutrino
and anti--neutrino disintegration of the deuteron at low energies
[26]. Thus, the NNJL model provides a theoretical foundation for a new
check of a value of the astrophysical factor for the solar proton
burning in terrestrial laboratories. At present the enhancement of the
astrophysical factor given by Eq.(\ref{label3.14}) does not contradict
to the available experimental data on the cross sections for the
disintegration of the deuteron by reactor anti--neutrinos [32]. The
theoretical cross sections fit the meanvalues of the experimentally
measured cross sections for the reactions $\bar{\nu}_{\rm e}$ + D
$\to$ $\bar{\nu}_{\rm e}$ + n + p and $\bar{\nu}_{\rm e}$ + D $\to$
e$^+$ + n + n [32] with an accuracy 1$\,\sigma$ and 1.32$\,\sigma$,
respectively [26].

The value of the astrophysical factor $S^{\rm KR}_{\rm pp}(0) =
(4.31\pm 0.35)\times 10^{-25}\,{\rm MeV\,b}$ having been recently
calculated by Kong and Ravndal [23] in the EFT agrees with the value
$S^{\rm ASM}_{\rm pp}(0) =(4.42^{+0.30}_{-0.08})\times 10^{-25}\,{\rm
MeV\,b}$ given by Eq.(\ref{label3.14}).  The astrophysical factor in
the Kong--Ravndal approach depends on the unknown counter--term that
seems to be very similar to our parameter $\bar{\xi}$. {\it Assuming
the counter--term to have a natural magnitude} [23], Kong and Ravndal
have obtained the value $S^{\rm KR}_{\rm pp}(0) = (4.31\pm 0.35)\times
10^{-25}\,{\rm MeV\,b}$. As has been stated by Kong and Ravndal [23]
the true value of the counter--term can be determined from precise
measurements of the cross sections for neutrino disintegration of the
deuteron at low energies. This is just that we have pointed out above
and earlier in Ref.[26] concerning the parameter $\bar{\xi}$.

For the discussion of the solar {\it hep} neutrino problem [11,37]
appeared due to recent experiments by SUPERKAMIOKANDE Collaboration
[16] we need to calculate in the NNJL model the astrophysical factor
for the {\it hep} reaction p + ${^3}{\rm He}$ $\to$ ${^4}{\rm He}$ +
e$^+$ + $\nu_{\rm e}$. The former demands, in turn, the extension of
the NNJL model by the inclusion of the light nuclei ${^3}{\rm He}$,
${^3}{\rm H}$ and ${^4}{\rm He}$. This work is in progress [24].

\section*{Acknowledgement}

\hspace{0.2in} One of the authors (A.N. Ivanov) thanks
Acad. G.T.Zatsepin and the participants of the seminar of the Lebedev
Physical Institute of Academy of Sciences, where this paper has been
reported, for fruitful discussions. 

\newpage

\noindent Table 1.  Solar neutrino data, $1\,{\rm SNU} =
10^{-36}\,{\rm events/(atoms\cdot s)}$. The error is defined as
$\sigma = \sqrt{({\rm stat.})^2 +({\rm syst.})^2}$
\vspace{0.2in}

\begin{tabular}{|c|c|c|  }\hline
\cline{2-3}Experiment & Data $\pm \,\sigma$ & Units\\ \hline HOMESTAKE
& & \\ $\nu_{\rm e}$ + ${^{37}}{\rm Cl}$ $\to$ e$^-$ + ${^{37}}{\rm
Ar}$&$ 2.56\pm 0.23$& SNU \\ $E_{\rm th} = 0.81\,{\rm MeV}$ & & \\
\hline SAGE & & \\ $\nu_{\rm e}$ + ${^{71}}{\rm Ga}$ $\to$ e$^-$ +
${^{71}}{\rm Ge}$ & $75.4\pm 7.6$ & SNU \\ $E_{\rm th} =
0.23\,{\rm MeV}$ & & \\ \hline GALLEX & & \\ $\nu_{\rm e}$ +
${^{71}}{\rm Ga}$ $\to$ e$^-$ + ${^{71}}{\rm Ge}$ &
$77.5\pm 7.7$ & SNU\\ $E_{\rm th} = 0.23\,{\rm MeV}$ & &\\ \hline GNO & & \\ $\nu_{\rm e}$ +
${^{71}}{\rm Ga}$ $\to$ e$^-$ + ${^{71}}{\rm Ge}$ &
$65.8\pm 10.5$ & SNU\\ $E_{\rm th} = 0.23\,{\rm MeV}$ & &\\ \hline GALLEX + GNO & & \\ $\nu_{\rm e}$ +
${^{71}}{\rm Ga}$ $\to$ e$^-$ + ${^{71}}{\rm Ge}$ &
$74.1\pm 6.8$ & SNU\\ $E_{\rm th} = 0.23\,{\rm MeV}$ & & \\
\hline KAMIOKANDE & & \\ $\nu$ + e$^-$ $\to$ $\nu$ + e$^-$ & $2.80\pm
0.38$ & $10^6\,{\rm cm^{-\,2}s^{-\,1}}$\\ $E_{\rm th} = 7.0\,{\rm
MeV}$ & & \\ \hline SUPERKAMIOKANDE& & \\ $\nu$ + e$^-$ $\to$ $\nu$ +
e$^-$ & $2.40 \pm 0.09 $ & $10^6\,{\rm cm^{-2}s^{-1}}$\\ $E_{\rm th} =
5.5\,{\rm MeV}$ & & \\
\hline
\end{tabular}\\
\vspace{0.2in}

\noindent Table 2.  Standard Solar Model predictions for the solar
neutrino fluxes normalized to the recommended value of the
astrophysical factor $S_{\rm pp}(0) = 4.00\times 10^{-25}\,{\rm
MeV\,b}$ [11].
\vspace{0.2in}

\begin{tabular}{|c|c|c|c|c|  }\hline
\cline{2-4} Source & Flux & Cl& Ga & SK
\\ 
& $(10^{10}\,{\rm cm^{-2}s^{-1}})$ & (SNU)& (SNU)& $(10^6\,{\rm cm^{-2}s^{-1}})$\\
\hline &   &   &  &\\ 
pp &  $5.94(1.00^{+0.01}_{-0.01})$  & 0.0 & 69.6 & \\ pep & $1.39\times 10^{-2}(1.00^{+0.01}_{-0.01})$ &
0.2& 2.8 &\\ $^7{\rm Be}$ & $4.80\times 10^{-1}(1.00^{+0.09}_{-0.09})$ & 1.15 & 34.4 & \\
$^{8}{\rm B}$ & $5.15\times 10^{-4}(1.00^{+0.19}_{-0.14})$ & 5.9 & 12.4 & $5.15^{+1.0}_{-0.7}$\\ $^{13}{\rm N}$ &
$6.05\times 10^{-2}(1.00^{+0.19}_{-0.13})$ & 0.1 & 3.7 & \\ $^{15}{\rm O}$ & $5.32\times
10^{-2}(1.00^{+0.22}_{-0.15})$ & 0.4 & 6.0& \\ \hline & & $7.7^{+1.2}_{-1.0}$ &
$129^{+8}_{-6}$ & $5.15^{+1.0}_{-0.7}$\\ \hline
\end{tabular}\\
\vspace{0.2in}

\newpage

\noindent Table 3. The NNJL model predictions for the solar neutrino
fluxes normalized to astrophysical factor $S_{\rm pp}(0) = 4.72\times
10^{-25}\,{\rm MeV\,b}$ caused by the non--trivial contribution of the
nucleon tensor current.

\vspace{0.2in}

\begin{tabular}{|c|c|c|c|c|  }\hline
\cline{2-4} Source & Flux & Cl& Ga & SK
\\ 
& $(10^{10}\,{\rm cm^{-2}s^{-1}})$ & (SNU)& (SNU)& $(10^6\,{\rm cm^{-2}s^{-1}})$\\
\hline pp & 6.10 & 0.0 & 71.49 & \\ pep & $1.48\times 10^{-2}$ &
0.21& 2.98 &\\ $^7{\rm Be}$ & $3.66\times 10^{-1}$ & 0.88 & 26.23 & \\
$^{8}{\rm B}$ & $2.69\times 10^{-4}$ & 3.08 & 6.46 & 2.69\\ $^{13}{\rm N}$ &
$3.13\times 10^{-2}$ & 0.05 & 1.91& \\ $^{15}{\rm O}$ & $2.56\times
10^{-2}$ & 0.19 & 2.89& \\ \hline & & $4.41$ &
$111.96$ & \\ \hline
\end{tabular}\\
\vspace{0.2in}

\noindent Table 4. The solar neutrino fluxes normalized to $S^{\rm
ASM}_{\rm pp}(0)= 1.18\,S_{\rm pp}(0) = 4.72\times 10^{-25}\,{\rm
MeV\,b}$. The theoretical values of experimentally measured neutrino
fluxes are calculated within a scenario of vacuum two--flavour
neutrino oscillations at $\sin^22\theta = 0.838$.  The error is
defined as $\sqrt{({\rm stat.})^2 +({\rm syst.})^2}$

\vspace{0.2in}

\begin{tabular}{|c|c|c|c|c|  }\hline
\cline{2-4} Source & Flux & Cl& Ga & SK \\ & $(10^{10}\,{\rm
cm^{-2}s^{-1}})$ & (SNU)& (SNU)& $(10^6\,{\rm cm^{-2}s^{-1}})$\\
\hline pp & $6.10$ & 0.0 & $41.54$ & \\ 
pep & $1.48\times 10^{-2}$ 
& $0.13$ &1.73 &\\ 
$^7{\rm Be}$ &
$3.66\times 10^{-1}$ & $0.50$ & $15.25$ & \\ 
$^{8}{\rm B}$ & $2.69\times
10^{-4}$ & $1.79$ & $3.74$ & $1.73$\\ 
$^{13}{\rm N}$ & $3.13\times 10^{-2}$
& $0.03$ & $1.11$ & \\ 
$^{15}{\rm O}$ & $2.56\times 10^{-2}$ & $0.11$ & $1.68$ &
\\ \hline & & $2.56^{+0.40}_{-0.33}$ & $65.05^{+4.03}_{-3.06}$ & $1.73^{+0.34}_{-0.24}$\\ 
\hline & & $2.56\pm 0.23$ & $74.1\pm 6.8$ & $2.40 \pm 0.09$\\ \hline
\end{tabular}\\

\noindent Table 5. The solar neutrino fluxes normalized to $S^{\rm
ASM}_{\rm pp}(0)= 4.42\times 10^{-25}\,{\rm MeV\,b}$. The theoretical
values of experimentally measured neutrino fluxes are calculated within
a scenario of vacuum two--flavour neutrino oscillations at
$\sin^22\theta = 0.780$.  The error is defined as $\sqrt{({\rm
stat.})^2 +({\rm syst.})^2}$

\vspace{0.2in}

\begin{tabular}{|c|c|c|c|c|  }\hline
\cline{2-4} Source & Flux & Cl& Ga & SK \\ & $(10^{10}\,{\rm
cm^{-2}s^{-1}})$ & (SNU)& (SNU)& $(10^6\,{\rm cm^{-2}s^{-1}})$\\
\hline pp & 6.07 & 0.0 & 43.40 & \\ pep & $1.46\times 10^{-2}$ & 0.13
&1.79 &\\ $^7{\rm Be}$ & $3.93\times 10^{-1}$ & 0.57 & 17.18 & \\
$^{8}{\rm B}$ & $3.18\times 10^{-4}$ & 2.22 & 4.67 & $2.12$\\
$^{13}{\rm N}$ & $3.71\times 10^{-2}$ & 0.05 & 1.39 & \\ $^{15}{\rm
O}$ & $3.09\times 10^{-2}$ & 0.14 & 2.13 & \\ \hline & & $3.11^{+0.49}_{-0.40}$ & $70.56^{+4.38}_{-3.32}$
& $2.12^{+0.41}_{-0.29}$\\ \hline & & $2.56\pm 0.23$ & $74.1\pm 6.8$ & $2.40 \pm 0.09$\\
\hline
\end{tabular}\\

\end{document}